\documentclass{article}

\usepackage{natbib}
\usepackage{hyperref}
\usepackage{graphicx}%
\usepackage{multirow}%
\usepackage{amsmath,amssymb,amsfonts}%
\usepackage{amsthm}%
\usepackage{mathrsfs}%
\usepackage{xcolor}%
\usepackage{textcomp}%
\usepackage{manyfoot}%
\usepackage{booktabs}%
\usepackage{algorithm}%
\usepackage{algorithmicx}%
\usepackage{algpseudocode}%
\usepackage{float}%

\newfloat{listing}{htbp}{lop}
\floatname{listing}{Listing}

\begin{document}

\title{Mining patterns in syntax trees to automate code reviews of student solutions for programming exercises}

\author{Charlotte {Van Petegem}, Kasper Demeyere, Rien Maertens,\\ Niko Strijbol, Bram {De Wever}, Bart Mesuere, Peter Dawyndt}

\maketitle

\section*{Abstract}
  In programming education, providing manual feedback is essential but labour-intensive, posing challenges in consistency and timeliness.
  We introduce ECHO, a machine learning method to automate the reuse of feedback in educational code reviews by analysing patterns in abstract syntax trees.
  This study investigates two primary questions: whether ECHO can predict feedback annotations to specific lines of student code based on previously added annotations by human reviewers (RQ1), and whether its training and prediction speeds are suitable for using ECHO for real-time feedback during live code reviews by human reviewers (RQ2).
  Our results, based on annotations from both automated linting tools and human reviewers, show that ECHO can accurately and quickly predict appropriate feedback annotations.
  Its efficiency in processing and its flexibility in adapting to feedback patterns can significantly reduce the time and effort required for manual feedback provisioning in educational settings.

\section*{Keywords}

Computer science education, Programming, Code review, Feedback provisioning, Automated assessment, Educational data mining

\section{Introduction}
\label{subsec:feedbackpredictionintro}
Feedback is a key factor in student learning~\citep{hattiePowerFeedback2007,blackAssessmentClassroomLearning1998}.
In programming education, many steps have been taken to give feedback using automated assessment systems~\citep{paivaAutomatedAssessmentComputer2022,ihantolaReviewRecentSystems2010,ala-mutkaSurveyAutomatedAssessment2005}.
These automated assessment systems give feedback on correctness, and can give some feedback on style and best practices through the use of linters.
However, they are generally unable to give the same high-level feedback on program design that an experienced programmer can give.
In many educational practices, automated assessment is therefore supplemented with manual feedback, especially when grading evaluations or exams~\citep{debuseEducatorsPerceptionsAutomated2008}.
This requires a significant time investment of teachers~\citep{tuckFeedbackgivingSocialPractice2012}.

As a result, many researchers have explored the use of AI to enhance giving feedback.
\citet{vittoriniAIBasedSystemFormative2021} used natural language processing to automate grading, and found that students who used the system during the semester were more likely to pass the course at the end of the semester.
\citet{leeSupportingStudentsGeneration2023} has used supervised learning with ensemble learning to enable students to perform peer and self-assessment.
In addition, \citet{berniusMachineLearningBased2022} introduced a framework based on clustering text segments in free-form textual exercises to reduce the grading workload.
\citet{strickrothSupportingSemiautomaticFeedback2023} attempt to solve this problem specifically for programming exercises by clustering submissions based on failed tests cases and compiler error messages.

\begin{figure}[htbp]
\centering
\includegraphics[width=.9\linewidth]{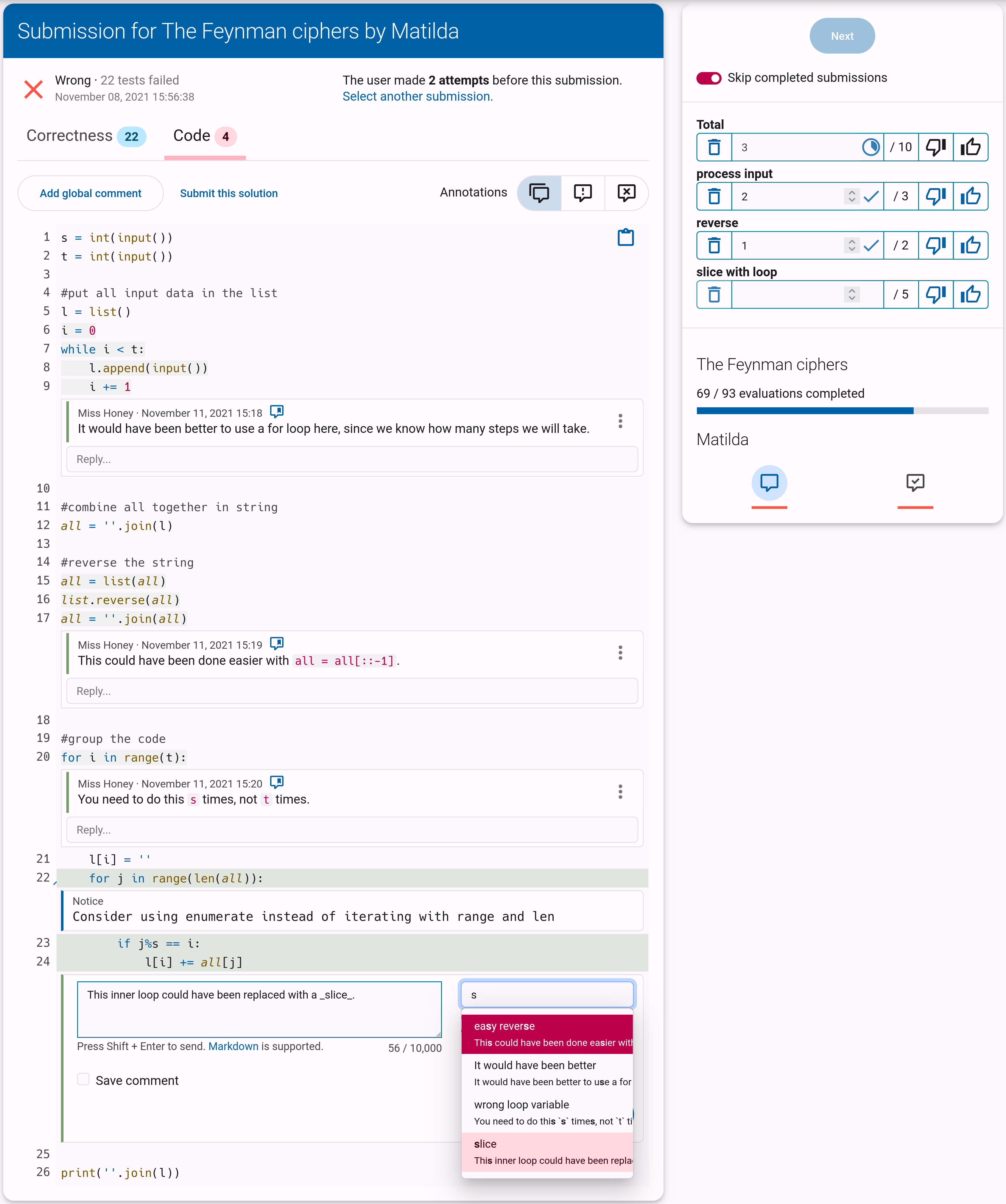}
\caption{\label{fig:feedbackintroductionreview}Assessment of a submitted solution in Dodona. An automated assessment has already been performed, with 22 failed test cases, as can be seen from the badge on the ``Correctness'' tab. An automated annotation left by Pylint can be seen on line 22. A teacher gives feedback on the code by adding inline annotations and scores the submission by filling out the exercise-specific scoring rubric. The teacher has just searched for a previously saved annotation so that they could reuse it. After manually assessing this submission, the teacher still has another 23 submissions to assess, as shown in the progress bar on the right.}
\end{figure}

The context of our work is the Dodona learning environment, developed at Ghent University~\citep{vanpetegemDodonaLearnCode2023}.
Dodona gives automated feedback on each submitted solution to programming exercises, but also has a module that allows teachers to give manual feedback on student submissions and assign scores.
The process of giving manual feedback on a submission to a programming exercise in Dodona is very similar to a code review, where errors or suggestions for improvements are annotated on the relevant line(s)~(Figure~\ref{fig:feedbackintroductionreview}).
In 2023 alone, 3\,663\,749 solutions were submitted to Dodona, of which 44\,012 were manually assessed.
During manual assessment, 22\,888 annotations were added to specific lines of code.

However, there is a crucial difference between traditional code reviews and those in an educational context: teachers often give feedback on numerous submissions to the same exercise.
Since students often make similar mistakes in their submissions to an exercise, it logically follows that teachers will repeatedly give the same or similar feedback on multiple student submissions.
To facilitate the reuse of feedback, Dodona allows teachers to save specific annotations for later search and retrieval.
In 2023, 777 annotations were saved by teachers on Dodona, which were reused a total of 7\,180 times.
The usage of this functionality has generated data that we can use in this study: annotations that are shared between code submissions and that occur on specific lines of those submissions.

In this manuscript we answer the following research questions:
(RQ1) Can previously added annotations be used to predict what annotations a reviewer is likely to add to a specific line of code during manual assessment of student-written code?
(RQ2) Additionally, can this be done so that both training of and predictions by the method are fast enough to use in live reviewing situations with human reviewers?

We present ECHO (Efficient Critique Harvesting and Ordering), a machine learning approach that aims to facilitate the reuse of previously given feedback.
We begin with a detailed explanation of the design of ECHO.
We then present and discuss the experimental results we obtained by testing ECHO on student submissions.
The dataset we used for this experiment is based on real Python code written by students during exams.
First, we test ECHO by predicting Pylint machine annotations.
Next, we use annotations left by human reviewers during manual assessment.

\section{Methodology}
\label{subsec:feedbackpredictionmethodology}
We consider predicting relevant annotations to be a ranking problem, which we solve by determining similarity between the lines of code where annotations are added.
The approach to determine this similarity is based on tree mining.
This is a data mining technique for extracting frequently occurring patterns from data that can be represented as trees~\citep{zakiEfficientlyMiningFrequent2005,asaiEfficientSubstructureDiscovery2004}.
Program code can be represented as an abstract syntax tree (AST), where the nodes of the tree represent the language constructs used in the program.
Recent work has demonstrated the efficacy of this approach in efficiently identifying frequent patterns in source code~\citep{phamMiningPatternsSource2019}.
In an educational context, these techniques have already been used to find patterns common to solutions that failed a given exercise~\citep{mensGoodBadUgly2021}.
Other work has demonstrated the potential of automatically generating unit tests from mined patterns~\citep{lienard2023extracting}.
We use tree mining to find commonalities between the lines of code where the same annotation has been added.

\begin{figure}[htbp]
\centering
\includegraphics[width=.9\linewidth]{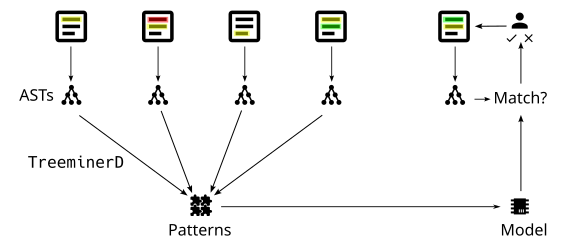}
\caption{\label{fig:feedbackmethodoverview}Overview of ECHO. Code of previously reviewed submissions is converted to its abstract syntax tree (AST) form. Instances of the same annotation have the same colour. For each annotation, the context of each instance is extracted and mined for patterns using the \texttt{TreeminerD} algorithm. These patterns are then weighted to form our model. When a reviewer wants to place an annotation on a line of the submissions they are currently reviewing, all previously given annotations are ranked based on the similarity determined for that line. The reviewer can then choose which annotation they want to place, with the aim of having the selected annotation at the top of in the ranking.}
\end{figure}

We begin with a general overview of ECHO (Figure~\ref{fig:feedbackmethodoverview}).
The first step is to use the tree-sitter library~\citep{brunsfeldTreesitterTreesitterV02024} to generate ASTs for each submission.
Using tree-sitter makes ECHO independent of the programming language used, since it presents an interface for generating syntax trees independent of the programming language.
The syntax trees are post-processed to include identifier names.
For each annotation, we identify all occurrences and extract a constrained AST context around the annotated line for each instance.
The resulting subtrees are then aggregated for each annotation.
If there are three or more subtrees, they are processed by the \texttt{TreeminerD} algorithm~\citep{zakiEfficientlyMiningFrequent2005}.
This yields a set of frequently occurring patterns specific to that annotation.
We then assign weights to these patterns based on their length and their frequency across the entire dataset of patterns for all annotations.
In addition to pattern mining, we also determine a set of unique nodes per forest of subtrees.
The result of these operations is our trained model.

The model can then be used to score how well an annotation matches a given code fragment.
In practice, the reviewer first selects a line of code in a given student's submission.
Next, the AST of the selected line and its surrounding context is generated.
For each annotation, each of its patterns is matched to the line, and a similarity score is calculated, given the previously determined weights.
The percentage of unique nodes which match in the current subtree is also taken into account.
These scores are used to rank the annotations, which are then displayed to the reviewer.
It is important to note that the reviewer remains in control of which annotation is used, if any.

We will now provide a more in-depth explanation of this process, with a particular emphasis on operational efficiency.
Speed is of the utmost importance throughout the model's lifecycle, from training to deployment in real-time reviewing contexts.
Given the continuous generation of training data during the review process, the model's training time must be optimized to avoid significant delays, ensuring that the model remains practical for live review situations.

\subsection{Training}
\label{subsubsec:feedbackpredictionsubtree}
\begin{figure}[htbp]
\centering
\includegraphics[width=0.5\linewidth]{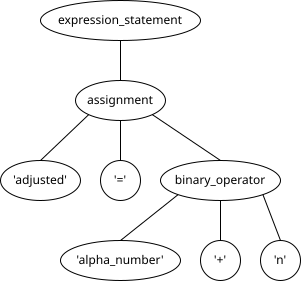}
\caption{\label{fig:feedbacksubtree}AST subtree corresponding to line 3 in Listing~\ref{lst:feedbacksubtreesample} as generated by tree-sitter.}
\end{figure}

The first step of ECHO is to extract a subtree for each instance of an annotation and then aggregate them per annotation.
Currently, the context around a line is extracted by taking all the AST nodes from that line.
For example, Figure~\ref{fig:feedbacksubtree} shows that the subtree extracted for the code on line 3 of Listing~\ref{lst:feedbacksubtreesample}.
Note that the context we extract here is very limited.
Previous iterations of ECHO considered all nodes that contained the relevant line (e.g. the function node for a line in a function), but these contexts proved too large to process in an acceptable time.

\begin{listing}[htbp]
\begin{verbatim}
def jump(alpha, n):
    alpha_number = ord(alpha)
    adjusted = alpha_number + n
    return chr(adjusted)
\end{verbatim}
\caption{\label{lst:feedbacksubtreesample}Example code that simply adds a number to the ASCII value of a character and converts it back to a character.}
\end{listing}

\begin{figure}[htbp]
\centering
\includegraphics[width=0.5\linewidth]{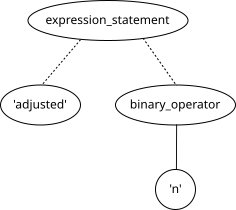}
\caption{\label{fig:feedbackpattern}Valid pattern for the tree in Figure~\ref{fig:feedbacksubtree}. Indirect ancestor-descendant relationships are marked with dashed lines.}
\end{figure}

After collecting subtrees for each annotation, ECHO mines patterns from these subtrees using \texttt{TreeminerD}~\citep{zakiEfficientlyMiningFrequent2005}: an algorithm for discovering frequently occurring patterns in datasets of rooted, ordered and labelled trees.
\texttt{TreeminerD} starts with a list of frequently occurring nodes, and then iteratively expands the frequently occurring patterns.
Patterns are embedded subtrees: the nodes in a pattern are a subset in the nodes of the tree, preserving the ancestor-descendant relationships and the left-to-right order of the nodes.
An example of a valid pattern for the tree in Figure~\ref{fig:feedbacksubtree} is shown in Figure~\ref{fig:feedbackpattern}.

In the \texttt{TreeminerD} algorithm, frequent means that the number of times the pattern occurs in all trees divided by the number of trees is greater than some predefined threshold.
This is called the \texttt{minimum support} parameter of the algorithm.

We use a custom implementation of the \texttt{TreeminerD} algorithm, to find patterns in the AST subtrees for each annotation.
Due to the exponential nature of the number of possible patterns in a tree, we only mine for patterns when there are at least three trees.

ECHO now has a set of patterns corresponding to each annotation.
However, some patterns are more informative that others.
So it assigns weights to the patterns it gets from \texttt{TreeminerD}.

Weights are assigned using two criteria.
The first criterion is the size of the pattern (i.e., the number of nodes in the pattern), since a pattern with twenty nodes is much more specific than a pattern with only one node.
The second criterion is the number of occurrences of a pattern across all annotations.
If the pattern sets for all annotations contain a particular pattern, it cannot be used reliably to determine which annotation should be predicted and is therefore given a lower weight.
Weights are calculated using the formula below.
\[\operatorname{weight}(pattern) = \frac{\operatorname{len}(pattern)}{\operatorname{\#occurences}(pattern)}\]

In addition to mining and weighting patterns, ECHO also determines a set of nodes that are unique to the subtrees of each annotation.
This is done by taking the union of the nodes of all subtrees for that annotation, and then removing from that set any nodes that occur in the subtrees of at least three other annotations.
This step does not require a minimum number of instances per annotation.

\subsection{Ranking}
\label{subsubsec:feedbackpredictionmatching}
Having completed the above steps, ECHO has trained its model.
To use the model, ECHO needs to know how to match patterns to subtrees.

To check whether a given pattern matches a given subtree, we iterate over all the nodes in the subtree.
At the same time, we also iterate over the nodes in the pattern.
During the iteration, we also store the current depth, both in the pattern and the subtree.
We also keep a stack to store (some of) the depths of the subtree.
If the current label in the subtree and the pattern are the same, we store the current subtree depth on the stack and move to the next node in the pattern.
Moving up in the tree is more complicated.
If the current depth and the depth of the last match (stored on the stack) are the same, we can move forwards in the pattern (and the subtree).
If not, we need to check that we are still in the embedded subtree, otherwise we need to reset our position in the pattern to the start.
Since subtrees can contain multiple instances of the same label, we also need to make sure that we can backtrack.
Listing~\ref{lst:feedbackmatchingpseudocode1} give the full pseudocode for this algorithm.

\begin{listing}[htbp]
\begin{verbatim}
start, p_i, pattern_depth, depth = 0
depth_stack, history = []

subtree_matches(subtree, pattern):
  result = find_in_subtree(subtree, subtree)
  while not result and history is not empty:
    to_explore, to_explore_subtree = history.pop()
    while not result and to_explore is not empty:
      start, depth, depth_stack, p_i = to_explore.pop()
      new_subtree = to_explore_subtree[start:]
      start = 0
      if pattern_length - p_i <= len(new_subtree) and \
          new_subtree is fully contained in pattern[p_i:]:
        result = find_in_subtree(subtree, new_subtree)
   return result

find_in_subtree(subtree, current_subtree):
  local_history = []
  for item in subtree:
    if item == -1:
      if depth_stack is not empty and depth - 1 == depth_stack.last:
        depth_stack.pop()
        if pattern [p_i] != -1:
          p_i = 0
          if depth_stack is empty:
            history.append((local_history, current_subtree[:i + 1])
            local_history = []
        else:
          p_i += 1
      depth -= 1
    else:
      if pattern[p_i] == item:
        local_history.append((start + i + 1, depth + 1, depth_stack, p_i))
        depth_stack.append(depth)
        p_i += 1
      depth += 1
    if p_i == pattern_length:
      return True
  if local_history is not empty:
    history.append((local_history, current_subtree))
  return False
\end{verbatim}
\caption{\label{lst:feedbackmatchingpseudocode1}Pseudocode for checking whether a pattern matches a subtree. Note that both the pattern and the subtree are stored in the encoding described by~\citet{zakiEfficientlyMiningFrequent2005}.}
\end{listing}

Checking whether a pattern matches a subtree is an operation that ECHO has to perform many times.
For some annotations there are many patterns, and all patterns of all annotations are checked.
An important optimization we added was to run the algorithm in Listing~\ref{lst:feedbackmatchingpseudocode1} only if the set of labels in the pattern is a subset of the labels in the subtree.

Given a model where we have weighted patterns for each annotation, and a method for matching patterns to subtrees, we can now put the two together to make a final ranking of the available annotations for a given line of code.
We calculate a match score for each annotation using the formula below.
\[ \operatorname{score}(annotation) = \frac{\displaystyle\sum_{pattern \atop \in\, patterns} \begin{cases} \operatorname{weight}(pattern) & pattern \text{ matches} \\ 0               & \text{otherwise} \end{cases}}{\operatorname{len}(patterns)} \]
ECHO then ranks the annotations by combining the score and the percentage of nodes in the set of unique nodes for that annotation.

\section{Results and discussion}
\label{subsec:feedbackpredictionresults}
As a dataset to validate ECHO, we used Python code written by students for programming exercises from (different) exams.
The dataset contains between 135 and 214 submissions per exercise.
Each submission for a particular exercise is by a different student.
We first split the datasets equally into a training set and a test set.
This simulates the midpoint of an assessment session for the exercise.
During testing, we let our model suggest annotations for each of the lines that had an actual annotation associated with them in the test set.
We evaluate where the correct annotation is ranked.
We only look at the top five to get a good idea of how useful the suggested ranking would be in practice: if an annotation is not in the top five, we would expect the reviewer to have to search for it manually, rather than selecting it directly from the suggested ranking.

We first ran Pylint\footnote{\url{https://www.pylint.org/}} (version 3.1.0) on the students' submissions.
Pylint is a static code analyser for Python that checks for errors and code smells, and enforces a standard programming style.
We used Pylint's machine annotations as our training and test data.
We test per exercise because that's our main use case for ECHO, but we also run a test that combines all submissions from all exercises.
An overview of some annotation statistics for the data generated by Pylint can be found in Table~\ref{tab:feedbackresultsdatasetpylint}.

\begin{table}[htbp]
\centering
\caption{\label{tab:feedbackresultsdatasetpylint}Statistics of Pylint annotations for the programming exercises used in the benchmark.}
\begin{tabular}{lrrrrr}
Exercise & subm. & ann. & inst. & max & avg\\[0pt]
\hline
A last goodbye\footnotemark & 135 & 25 & 189 & 29 & 7.56\\[0pt]
Symbolic\footnotemark & 141 & 28 & 277 & 66 & 9.89\\[0pt]
Narcissus\footnotemark cipher & 144 & 29 & 148 & 24 & 5.10\\[0pt]
Cocktail bar\footnotemark & 211 & 31 & 162 & 29 & 5.23\\[0pt]
Anthropomorphic emoji\footnotemark & 214 & 24 & 144 & 40 & 6.00\\[0pt]
Hermit\footnotemark & 194 & 82 & 388 & 59 & 6.80\\[0pt]
Combined & 1039 & 82 & 1479 & 196 & 18.04\\[0pt]
\end{tabular}
\end{table}
\footnotetext[2]{\url{https://dodona.be/en/activities/505886137/}}
\footnotetext[3]{\url{https://dodona.be/en/activities/933265977/}}
\footnotetext[4]{\url{https://dodona.be/en/activities/1730686412/}}
\footnotetext[5]{\url{https://dodona.be/en/activities/1875043169/}}
\footnotetext[6]{\url{https://dodona.be/en/activities/2046492002/}}
\footnotetext[7]{\url{https://dodona.be/en/activities/2146239081/}}

In a second experiment, we used the manual annotations left by human reviewers on student code in Dodona.
Exercises were reviewed by different people, but all submissions for a specific exercise were reviewed by the same person.
The reviewers were not aware of ECHO at the time they reviewed the submissions.
In this case there is no combined test as the set of annotations used is different for each exercise.

We distinguish between these two sources of annotations because we expect Pylint to be more consistent in both when it places an instance of an annotation and also where it places the instance.
Most linting annotations are detected by explicit pattern matching in the AST, so we expect the implicit pattern matching to work fairly well.
However, we want to skip this explicit pattern matching for manual annotations because of the time it takes to compile them and the fact that annotations are often specific to a particular exercise and reviewer.
Therefore, we also test on manual annotations.
Manual annotations are expected to be more inconsistent because reviewers may miss a problem in one student's code that they have annotated in another student's code, or they may not place instances of a particular annotation in consistent locations.
The method by which human reviewers place an annotation is also much more implicit than Pylint's pattern matching.

The reviewed programming exercises have between 55 and 469 instances of manual annotations.
The number of distinct annotations varies between 7 and 34 per exercise.
Table~\ref{tab:feedbackresultsdataset} gives an overview of some of characteristics of the dataset.
Timings mentioned in this section were measured on a 2022 Dell laptop with a 3GHz Intel quad-core processor and 32 GB of RAM.

\begin{table}[htbp]
\centering
\caption{\label{tab:feedbackresultsdataset}Statistics of manually added annotations for the programming exercises used in the benchmark.}
\begin{tabular}{lrrrrr}
Exercise & subm. & ann. & inst. & max & avg\\[0pt]
\hline
A last goodbye & 135 & 34 & 334 & 92 & 9.82\\[0pt]
Symbolic & 141 & 7 & 55 & 25 & 7.85\\[0pt]
Narcissus cipher & 144 & 17 & 193 & 55 & 11.35\\[0pt]
Cocktail bar & 211 & 15 & 469 & 231 & 31.27\\[0pt]
Anthropomorphic emoji & 214 & 27 & 322 & 39 & 11.93\\[0pt]
Hermit & 194 & 32 & 215 & 27 & 6.71\\[0pt]
\end{tabular}
\end{table}

\subsection{Machine annotations (Pylint)}
\label{subsubsec:feedbackpredictionresultspylint}
\begin{figure}[htbp]
\centering
\includegraphics[width=.9\linewidth]{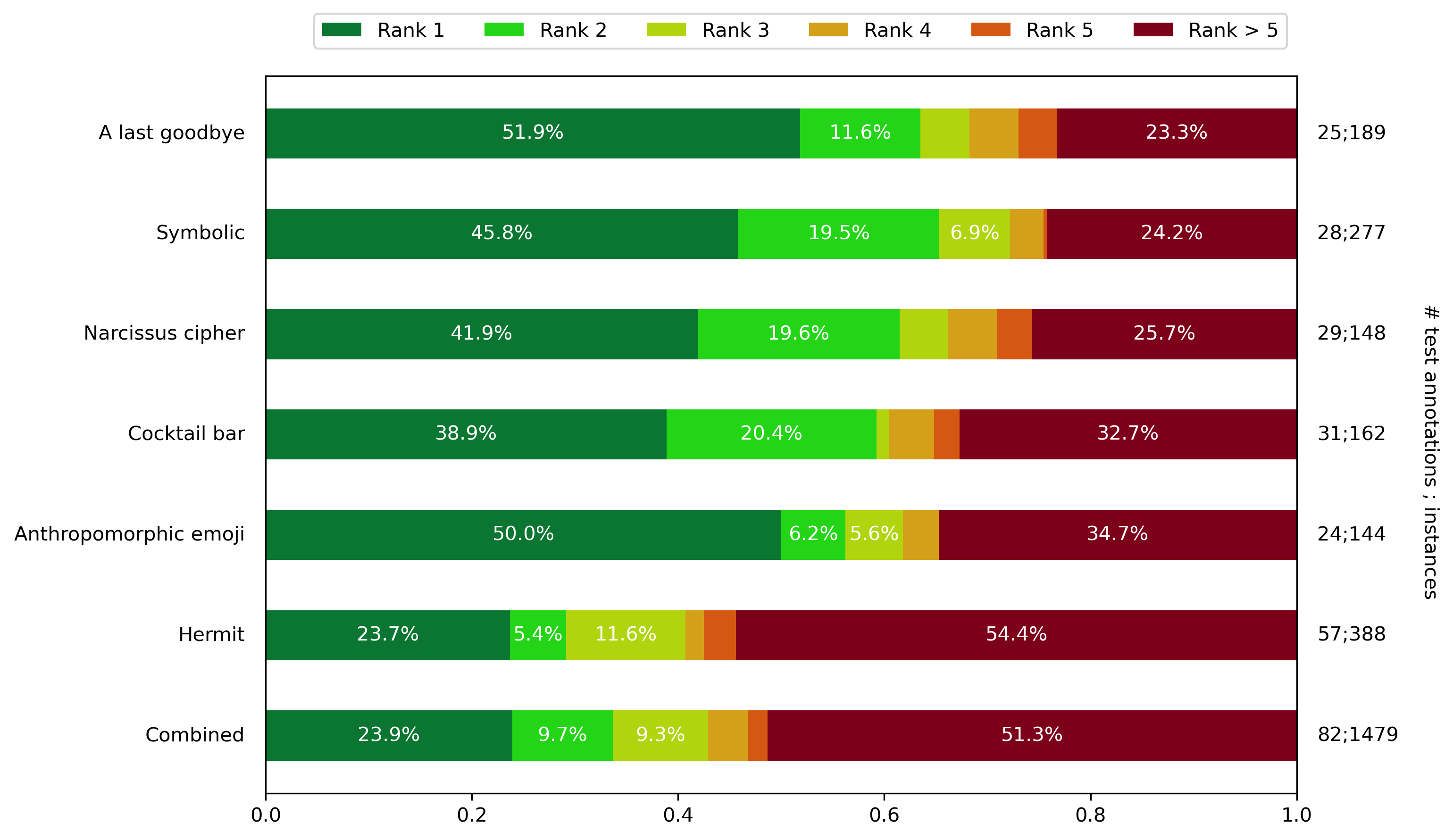}
\caption{\label{fig:feedbackpredictionpylintglobal}Prediction accuracy for suggesting instances of Pylint annotations where training and test  data are equally split. The numbers on the right are the total number of annotations and instances respectively. The ``Combined'' test evaluated ECHO on the entire set of submissions for all exercises.}
\end{figure}

We will first discuss the results for the Pylint annotations.
During the experiment, a few Pylint annotations that are not related to the structure of the code were omitted to avoid distorting the results.
These are ``line too long'', ``trailing whitespace'', ``trailing newlines'', ``missing module docstring'', ``missing class docstring'', and ``missing function docstring''.
Depending on the exercise, the actual annotation is ranked among the top five annotations in 45\% to 77\% of all test instances (Figure~\ref{fig:feedbackpredictionpylintglobal}).
The annotation is even ranked first for 23\% to 52\% of all test instances.
Interestingly, the method performs worse when the instances for all exercises are combined.
This highlights the fact that ECHO is most useful in the context of reviewing similar code many times.
For the submissions and instances in the training set, training took between 70 and 245 milliseconds to process all submissions and instances for an exercise.
The entire test phase took between 30 and 180 milliseconds per exercise.
Individual predictions never exceed 15 milliseconds.

\begin{figure}[htbp]
\centering
\includegraphics[width=.9\linewidth]{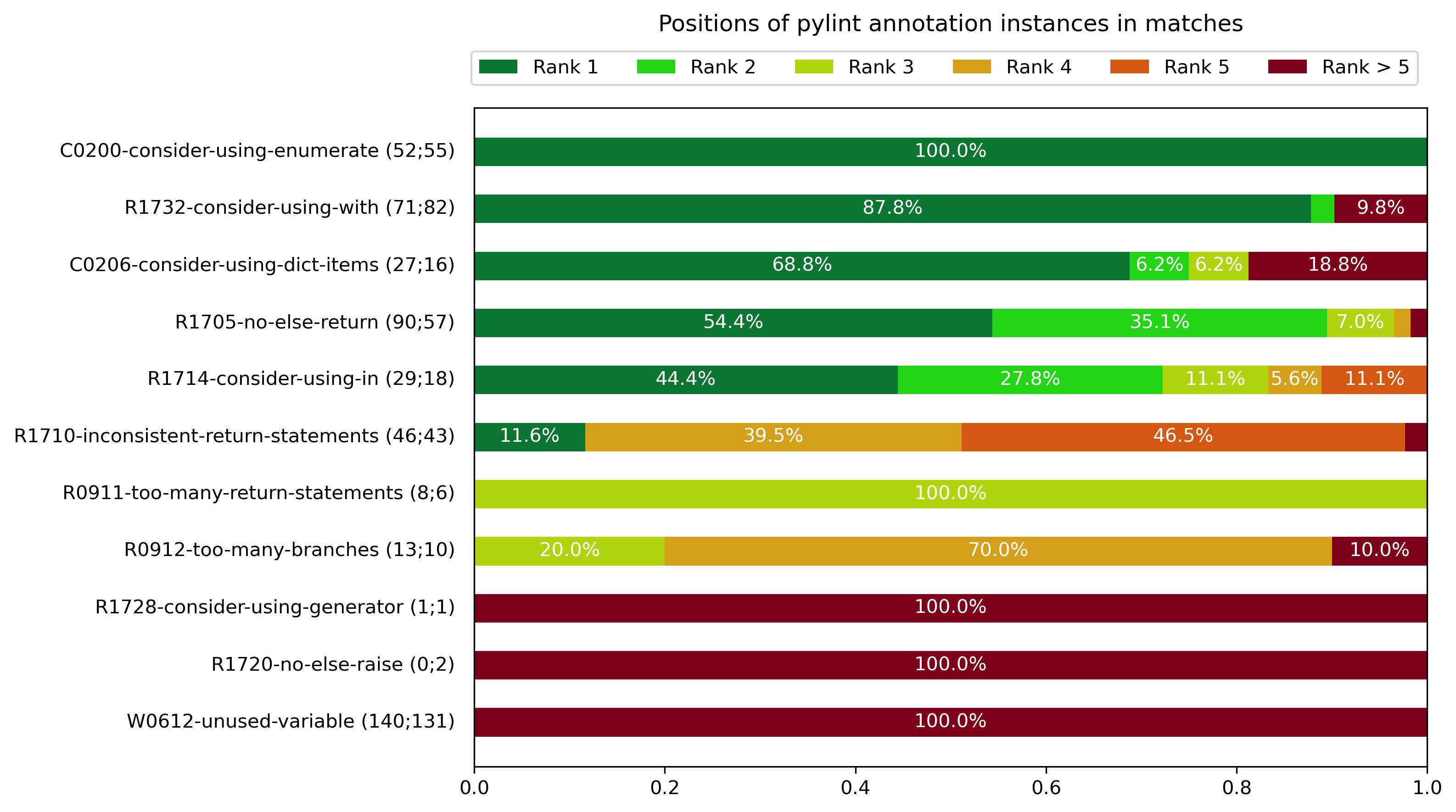}
\caption{\label{fig:feedbackpredictionpylintmessages}Prediction accuracy for a selection of Pylint machine annotations where training and test  data are equally split. Each line corresponds to a Pylint annotation, with the number of instances in the training and test sets given in parentheses after the annotation name.}
\end{figure}

We have selected some interesting annotations for further inspection, some of which perform very well, and some of which perform less well (Figure~\ref{fig:feedbackpredictionpylintmessages}).
We chose these specific annotations to demonstrate interesting behaviours exhibited by ECHO.
The differences in performance can be explained by the content of the annotation and the underlying patterns that Pylint is looking for.
For example, the ``unused variable''\footnote{\url{https://pylint.pycqa.org/en/latest/user\_guide/messages/warning/unused-variable.html}} annotation performs poorly.
This can be explained by the fact that we do not feed \texttt{TreeminerD} with enough context to find predictive patterns for this Pylint annotation.
There are also annotations that can't be predicted at all, because no patterns are found.

Other annotations, such as ``consider using with''\footnote{\url{https://pylint.pycqa.org/en/latest/user\_guide/messages/refactor/consider-using-with.html}}, work very well.
For these annotations, \texttt{TreeminerD} does have enough context to automatically determine the underlying patterns.
The number of instances of an annotation in the training set also has an effect.
Annotations with few instances are generally predicted worse than those with many instances.

\subsection{Human annotations}
\label{subsubsec:feedbackpredictionresultsrealworld}
For the annotations added by human reviewers, we used two different scenarios to evaluate ECHO.
In addition to using the same 50/50 split between training and test data as for the Pylint data, we also simulated how a human reviewer would use ECHO in practice by gradually increasing the training set and decreasing the test set as the reviewer progresses through the submissions during the assessment.
At the start of the assessment, no annotations are available and the first instance of an annotation that applies to a reviewed submission cannot be predicted.
As more submissions are reviewed and more instances of annotations are placed on those submissions, the training set for modelling predictions on the next submission under review gradually grows.

\begin{figure}[htbp]
\centering
\includegraphics[width=.9\linewidth]{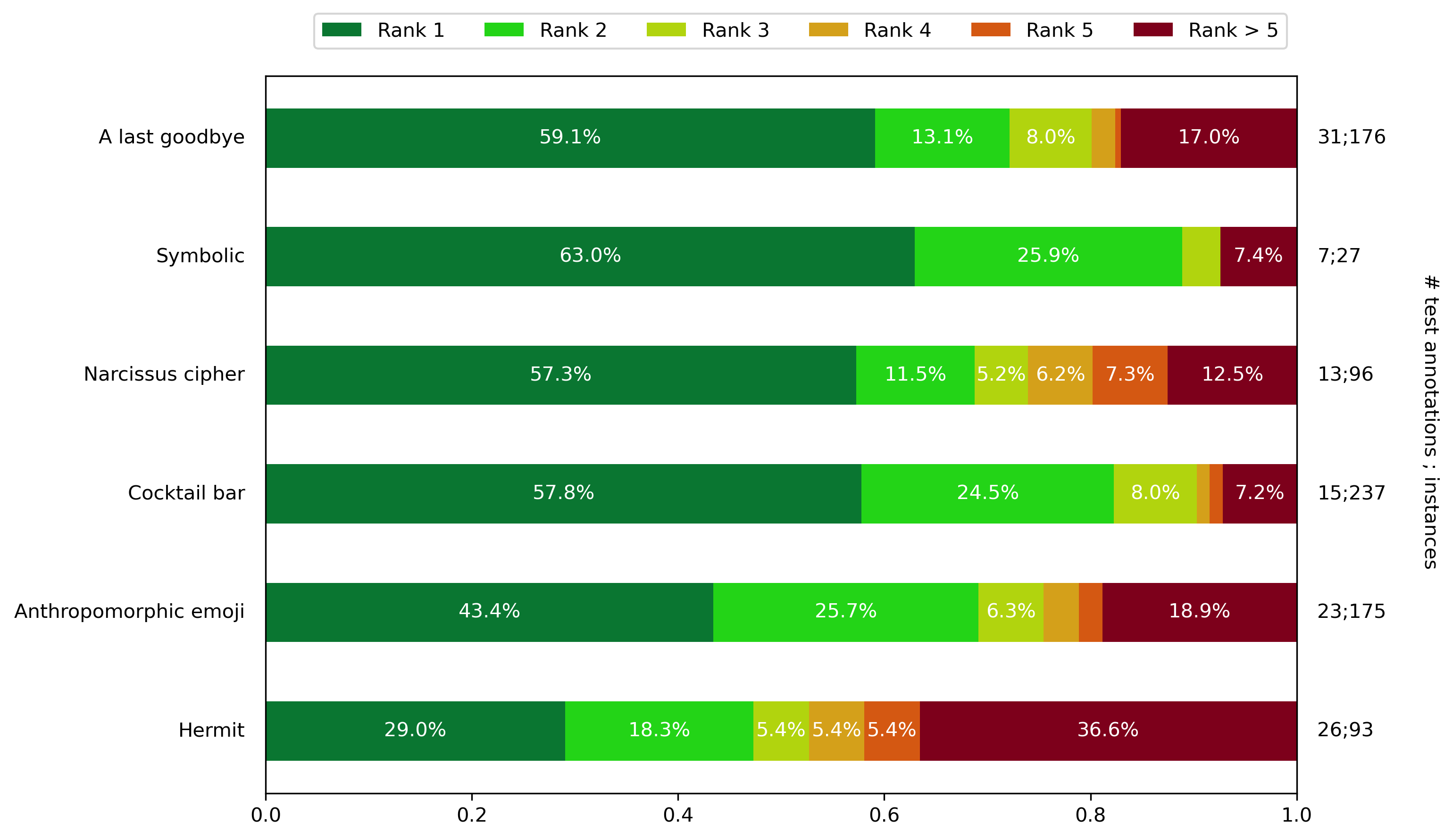}
\caption{\label{fig:feedbackpredictionrealworldglobal}Prediction accuracy for suggesting instances of annotations by human reviewers where training and test  data are equally split. The numbers on the right are the total number of annotations and instances respectively.}
\end{figure}

If we split the submissions and the corresponding annotations of a human reviewer equally into a training and a test set, the prediction accuracy is similar or even slightly better compared to the Pylint annotations (Figure~\ref{fig:feedbackpredictionrealworldglobal}).
The number of instances where the true annotation is ranked first is generally higher (between 29\% and 63\% depending on the exercise), and the number of instances where it is ranked in the top five is between 63\% and 93\% depending on the exercise.

In this experiment, training took between 67 milliseconds and 22.4 seconds per exercise.
The entire test phase took between 49 milliseconds and 27 seconds, depending on the exercise.
These evaluations were run on the same hardware as those for the machine annotations.
For one prediction, the average time ranged from 0.1 milliseconds to 150 milliseconds and the maxima from 0.5 milliseconds to 2.8 seconds.
The explanation for these wide ranges remains the same as for the Pylint predictions: it all depends on the number of patterns found.

These results show that we can predict reuse with a fairly high accuracy at the midpoint of a review session for a programming exercise.
The accuracy depends on the number of instances per annotation and the consistency of the reviewer.
Looking at the underlying data, we can also see that short, atomic messages can be predicted very well, as suggested by~\citet{moonsAtomicReusableFeedback2022}.
We will now look at the longitudinal prediction accuracy of ECHO, to test how accuracy evolves over the course of a review session.

For the next experiment, we introduce two specific categories of negative prediction results, namely ``No training instances'' and ``No patterns''.
``No training instances'' means that the annotation corresponding to the true instance had no instances in the training set, and therefore could never have been predicted.
``No patterns'' means that \texttt{TreeminerD} was unable to find any frequent patterns for the set of subtrees extracted from the annotation instances and there were also no nodes unique to this set of subtrees in the entire set of subtrees.
This could be because the collection of subtrees is too diverse to have common patterns.

\begin{figure}[htbp]
\centering
\includegraphics[width=.9\linewidth]{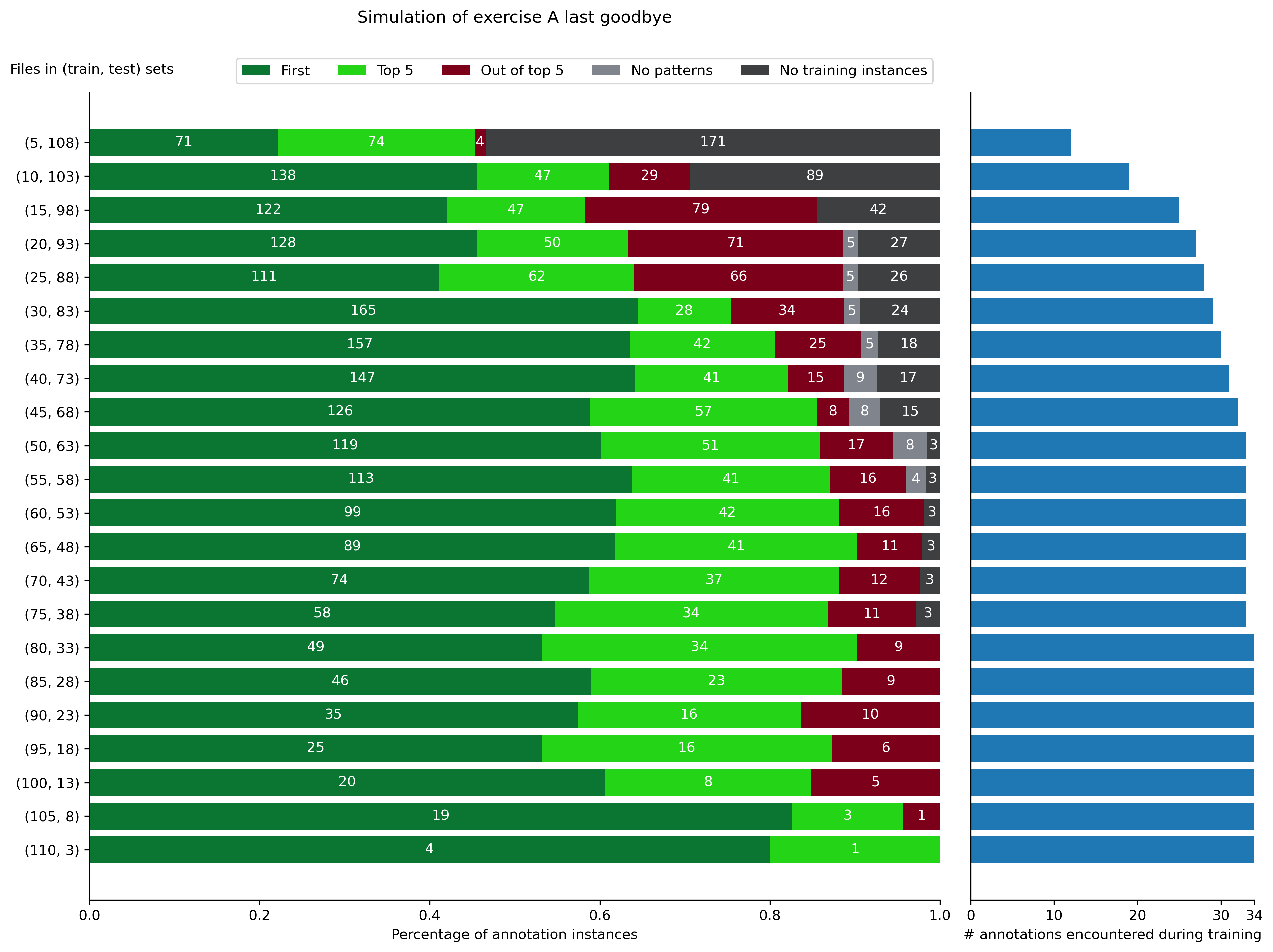}
\caption{\label{fig:feedbackpredictionrealworldsimulation1}Progression of the prediction accuracy for the ``A last goodbye'' exercise over the course of the review process. Predictions for instances whose annotation had no instances in the training set are classified as ``No training instances''. Predictions for instances whose annotation had no corresponding patterns in the model learned from the training set are classified as ``No patterns''. The graph on the right shows the number of annotations present with at least one instance in the training set.}
\end{figure}

\begin{figure}[htbp]
\centering
\includegraphics[width=.9\linewidth]{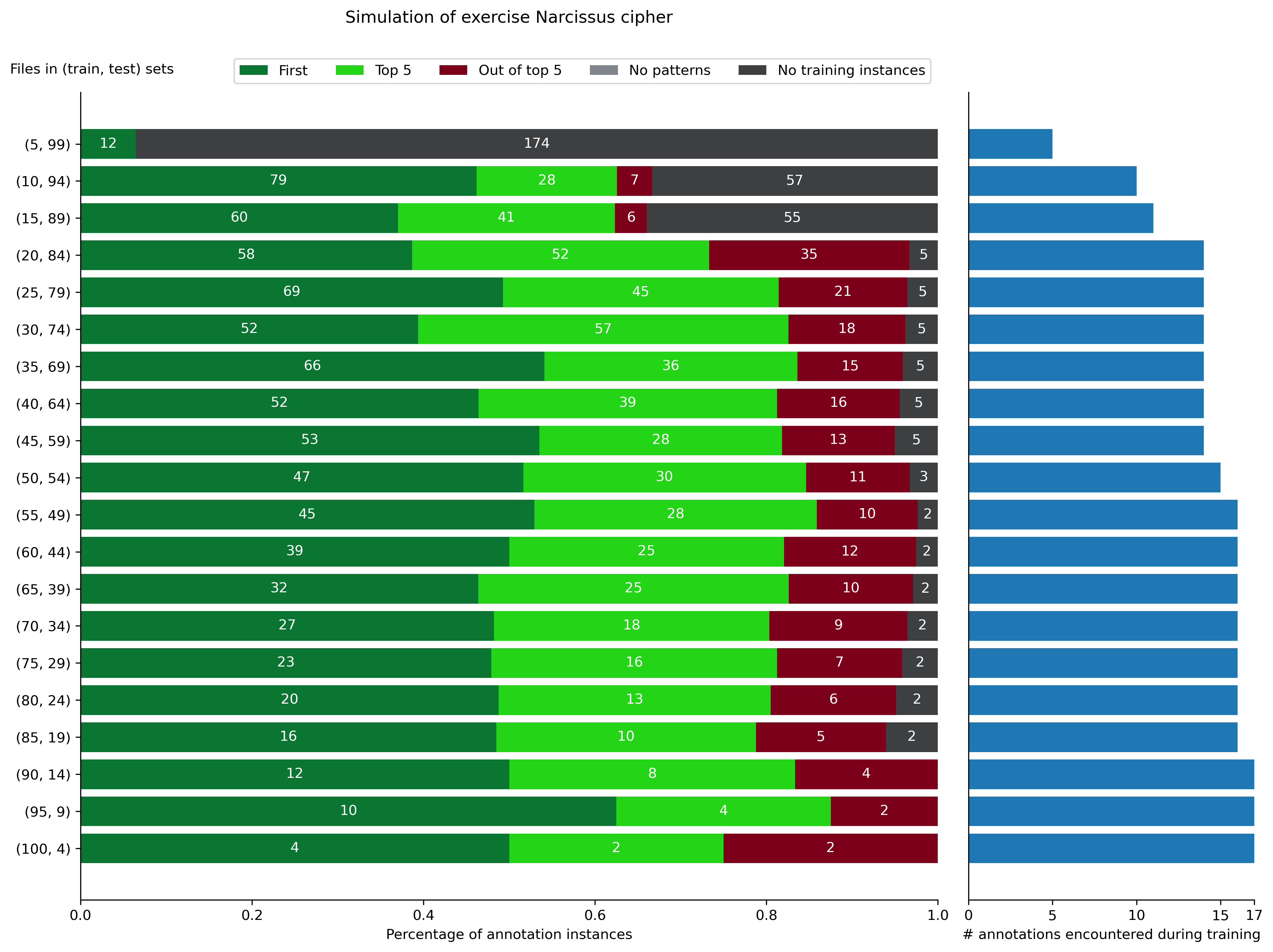}
\caption{\label{fig:feedbackpredictionrealworldsimulation2}Progression of the prediction accuracy for the ``Narcissus cipher'' exercise over the course of the review process. Predictions for instances whose annotation had no instances in the training set are classified as ``No training instances''. Predictions for instances whose annotation had no corresponding patterns in the model learned from the training set are classified as ``No patterns''. The graph on the right shows the number of annotations present with at least one instance in the training set.}
\end{figure}

\begin{figure}[htbp]
\centering
\includegraphics[width=.9\linewidth]{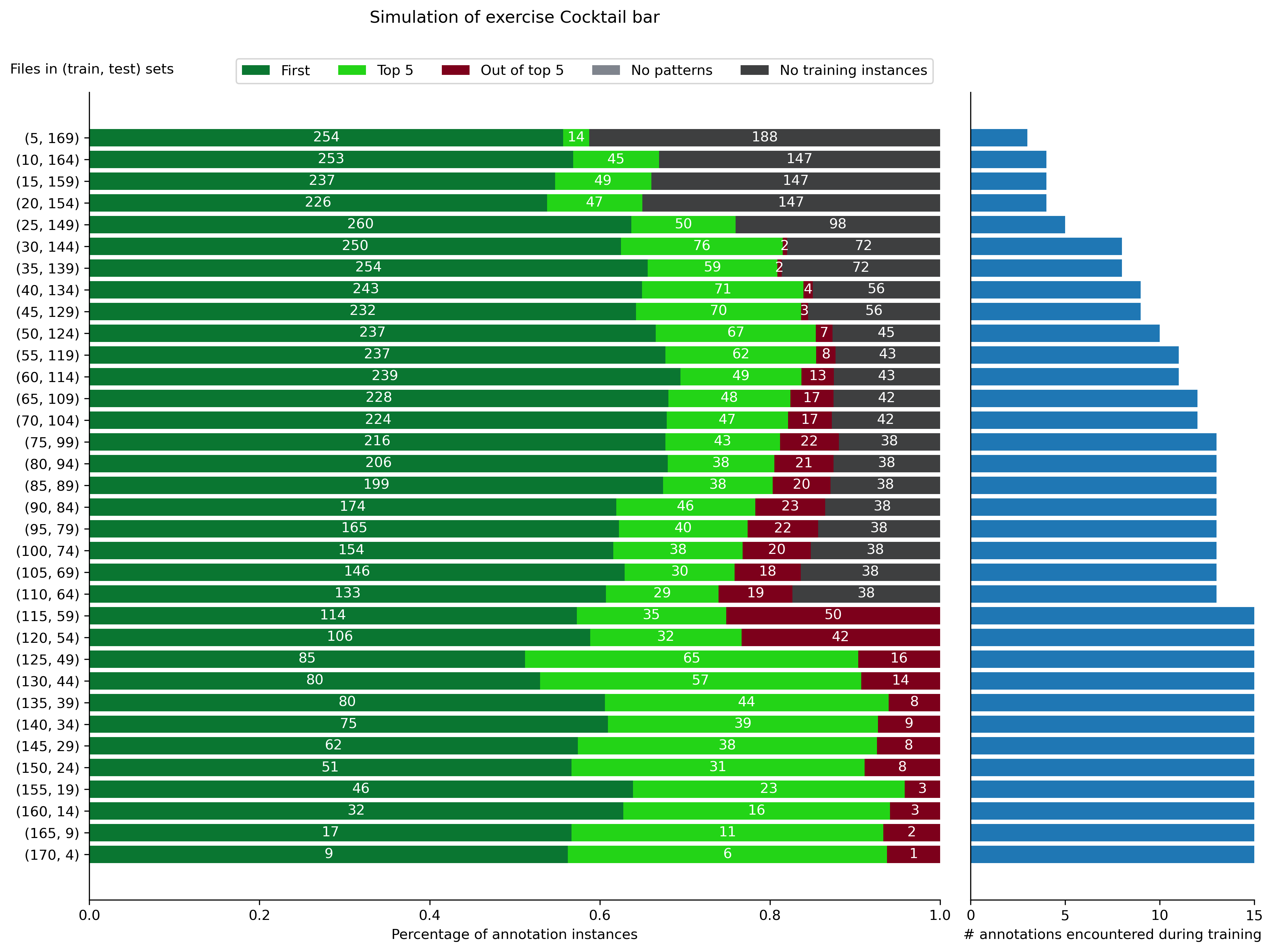}
\caption{\label{fig:feedbackpredictionrealworldsimulation3}Progression of the prediction accuracy for the ``Cocktail bar'' exercise over the course of the review process. Predictions for instances whose annotation had no instances in the training set are classified as ``No training instances''. Predictions for instances whose annotation had no corresponding patterns in the model learned from the training set are classified as ``No patterns''. The graph on the right shows the number of annotations present with at least one instance in the training set.}
\end{figure}

\begin{figure}[htbp]
\centering
\includegraphics[width=.9\linewidth]{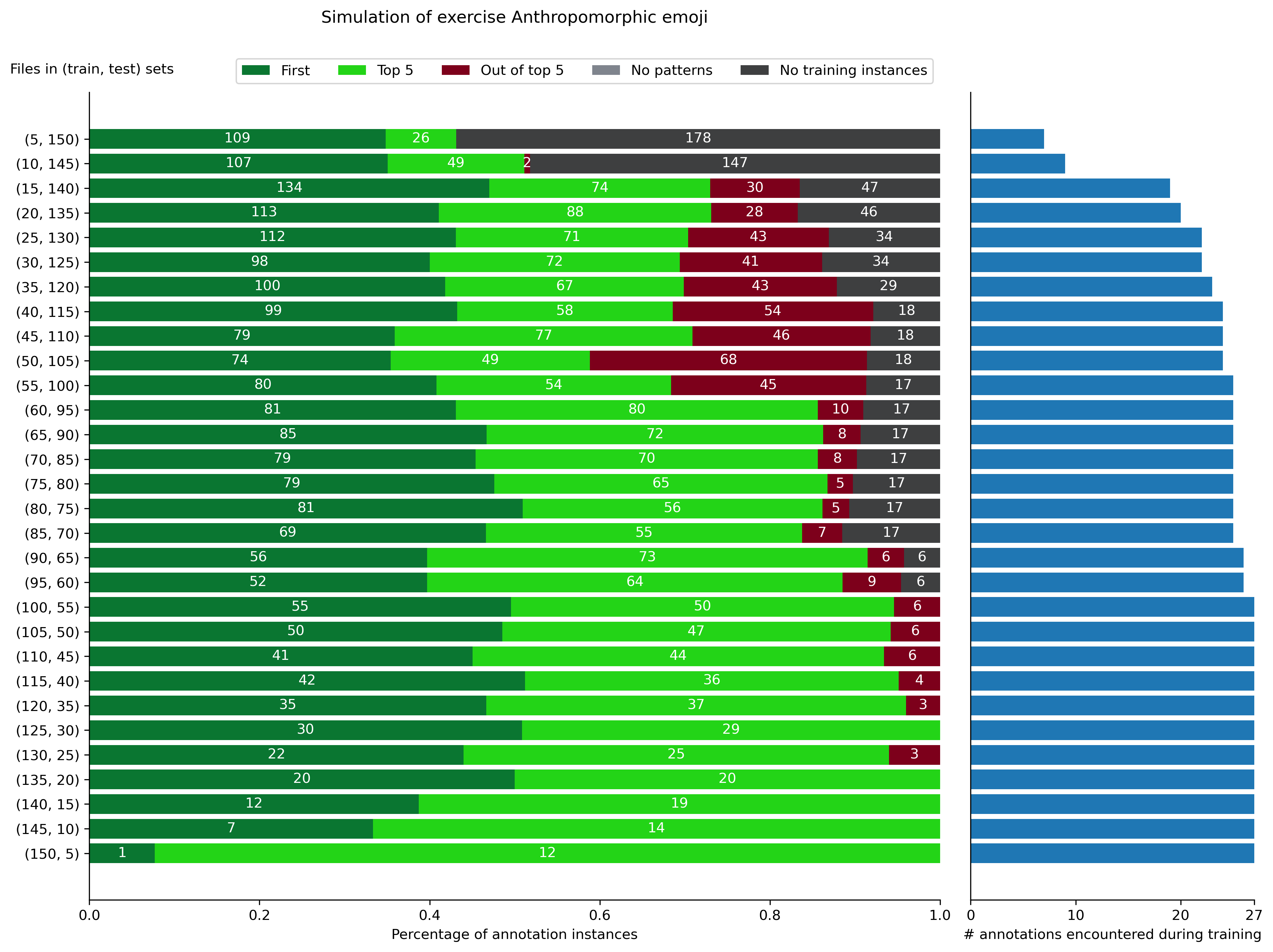}
\caption{\label{fig:feedbackpredictionrealworldsimulation4}Progression of the prediction accuracy for the ``Anthropomorphic emoji'' exercise over the course of the review process. Predictions for instances whose annotation had no instances in the training set are classified as ``No training instances''. Predictions for instances whose annotation had no corresponding patterns in the model learned from the training set are classified as ``No patterns''. The graph on the right shows the number of annotations present with at least one instance in the training set.}
\end{figure}

\begin{figure}[htbp]
\centering
\includegraphics[width=.9\linewidth]{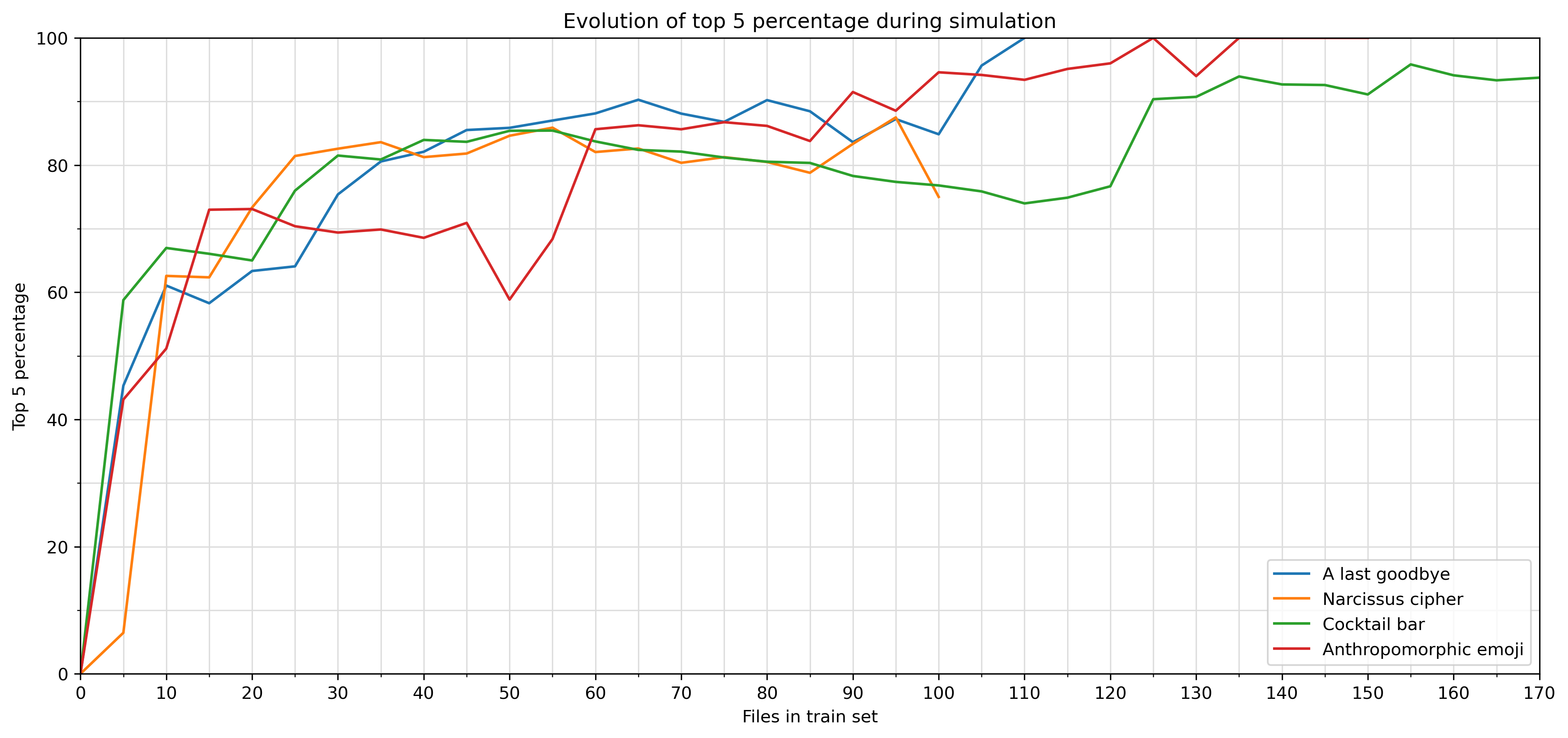}
\caption{\label{fig:feedbackpredictionrealworldevolution}Evolution of the percentage of suggestions that are ranked in the top~5. The percentages are fairly stable after 20 to 30 submissions have been reviewed.}
\end{figure}

Figures~\ref{fig:feedbackpredictionrealworldsimulation1},~\ref{fig:feedbackpredictionrealworldsimulation2},~\ref{fig:feedbackpredictionrealworldsimulation3}~and~\ref{fig:feedbackpredictionrealworldsimulation4} show the results of this experiment for four of the programming exercises used in the previous experiments.
The ``Symbolic'' exercise was excluded due to its low number of unique annotations, while the ``Hermit'' exercise was excluded due to its poor performance in the previous experiment.
We also excluded submissions that received no annotations during the human review process, which explains the lower number of submissions compared to the numbers in Table~\ref{tab:feedbackresultsdataset}.
This experiment shows that while the review process requires some time to build up before sufficient training instances are available, once a critical mass of training instances is reached, the accuracy for suggesting new instances of annotations reaches its maximum predictive power.
This critical mass is reached after about 20 to 30 submissions reviewed, which is quite early in the review process~(Figure~\ref{fig:feedbackpredictionrealworldevolution}).
This means that a lot of time could be saved during the review process when ECHO is integrated into an online learning environment.
The point at which the critical mass is reached will of course depend on the nature of the exercises and the consistency of the reviewer.

As mentioned above, we are working with a slightly inconsistent dataset when using annotations from human reviewers.
They will sometimes miss an instance of an annotation, place it inconsistently, or unnecessarily create duplicate annotations.
If ECHO is used in practice, the predictions may be even better, as the knowledge of its existence may further motivate reviewers to be more consistent in their reviews.
The programming exercises were also reviewed by different people, which may also explain the differences in prediction accuracy between the exercises.

\begin{figure}[htbp]
\centering
\includegraphics[width=.9\linewidth]{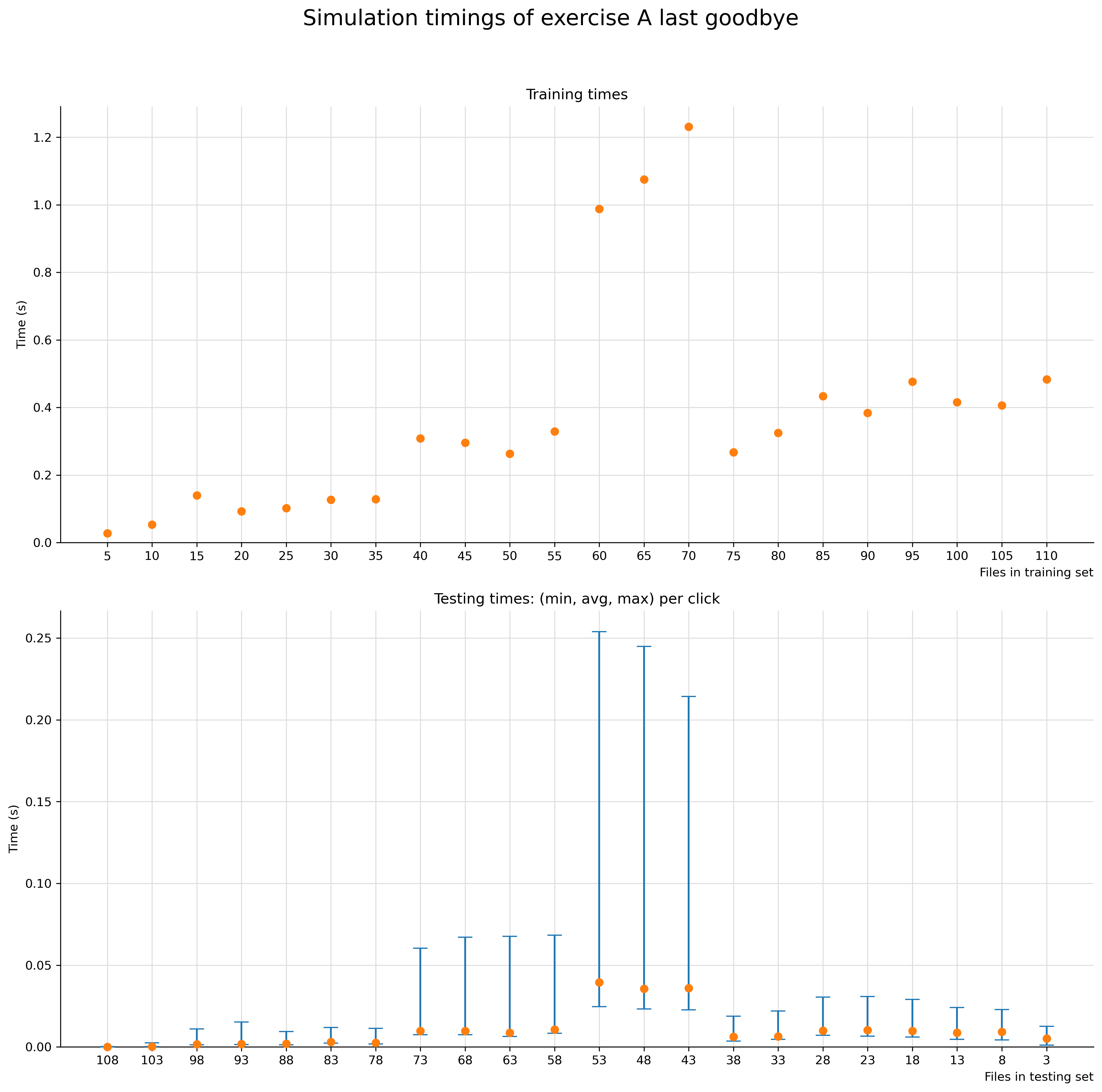}
\caption{\label{fig:feedbackpredictionrealworldtimings1}Time needed for training and testing during the entire review process for the exercise ``A last goodbye''. Top: training time. Bottom: average (orange dot) and range (blue line) of time needed to predict a single instance.}
\end{figure}

\begin{figure}[htbp]
\centering
\includegraphics[width=.9\linewidth]{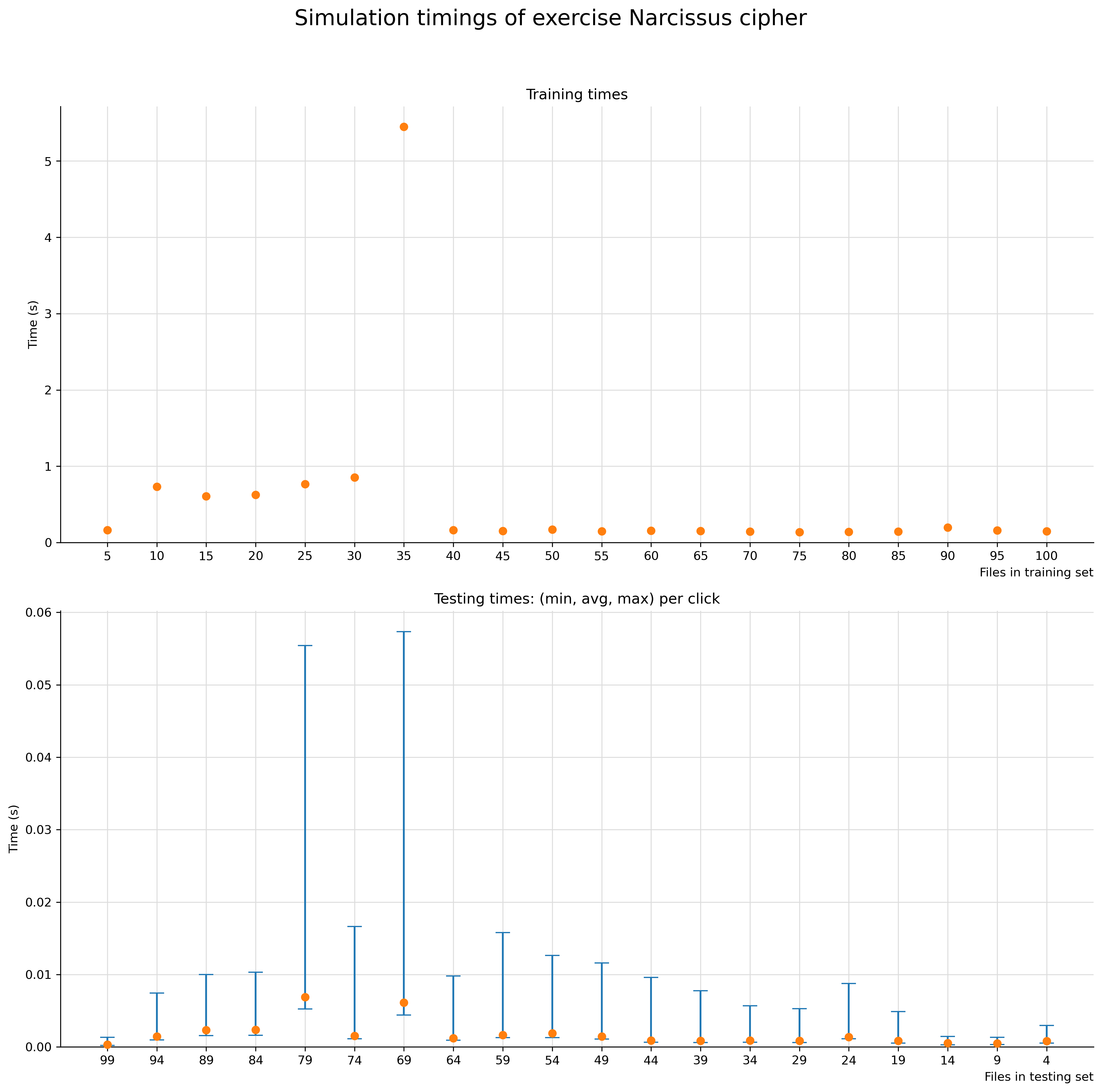}
\caption{\label{fig:feedbackpredictionrealworldtimings2}Time needed for training and testing during the entire review process for the exercise ``Narcissus cipher''. Top: training time. Bottom: average (orange dot) and range (blue line) of time needed to predict a single instance.}
\end{figure}

\begin{figure}[htbp]
\centering
\includegraphics[width=.9\linewidth]{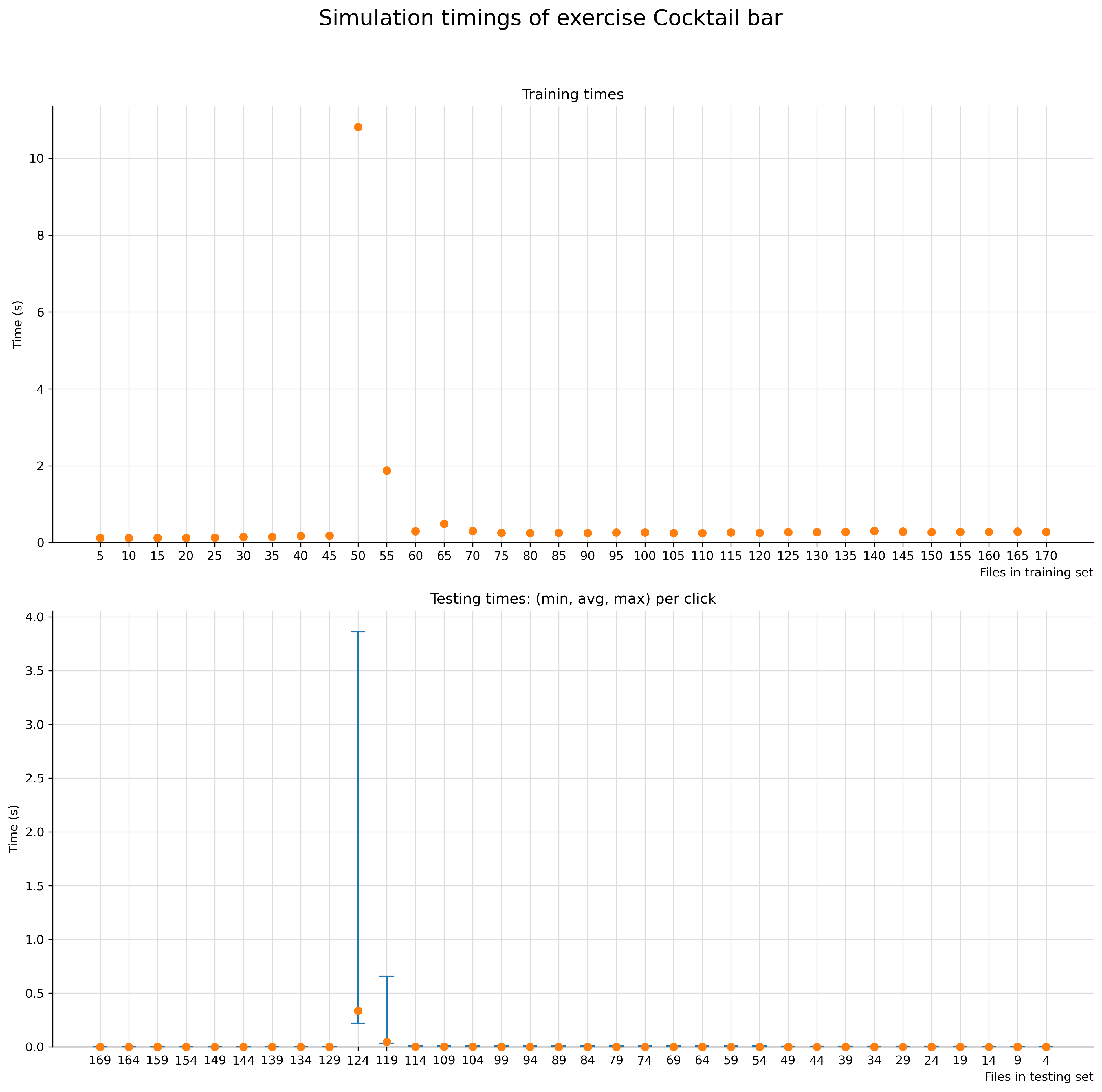}
\caption{\label{fig:feedbackpredictionrealworldtimings3}Time needed for training and testing during the entire review process for the exercise ``Cocktail bar''. Top: training time. Bottom: average (orange dot) and range (blue line) of time needed to predict a single instance.}
\end{figure}

\begin{figure}[htbp]
\centering
\includegraphics[width=.9\linewidth]{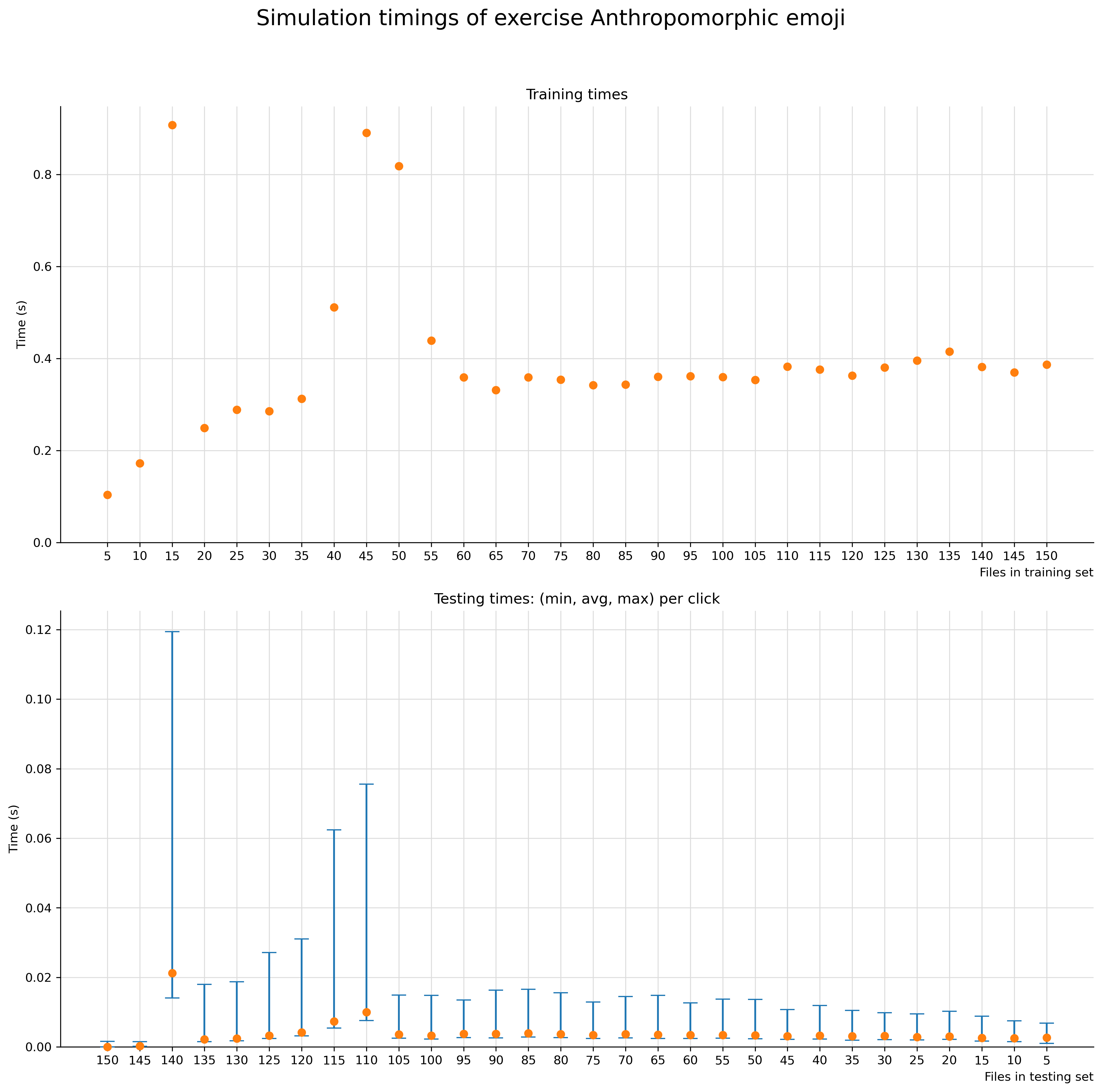}
\caption{\label{fig:feedbackpredictionrealworldtimings4}Time needed for training and testing during the entire review process for the exercise ``Anthropomorphic emoji''. Top: training time. Bottom: average (orange dot) and range (blue line) of time needed to predict a single instance.}
\end{figure}

To evaluate the performance of ECHO for these experiments, we measure the training times, and the times required for each prediction.
This corresponds to a reviewer wanting to add an annotation to a line in practice.
Figures~\ref{fig:feedbackpredictionrealworldtimings1},~\ref{fig:feedbackpredictionrealworldtimings2},~\ref{fig:feedbackpredictionrealworldtimings3},~and~\ref{fig:feedbackpredictionrealworldtimings4} show the performance of running these experiments.
As in the previous experiments, we can see that there is a considerable difference between the exercises.
However, the training time only exceeds one second in a few cases and remains well below that in most cases.
The prediction times are mostly below 50 milliseconds, except for a few outliers.
The average prediction time never exceeds 500 milliseconds.

The timings show that although there are some outliers, predictions can be made fast enough to make this an interactive system.
The outliers also correspond to higher training times, indicating that this is mainly caused by a high number of underlying patterns for some annotations.
Currently this process is also parallelized over the files, but in practice, the process could be parallelized over the patterns, which would speed up the prediction even more.
Note that the training time may also decrease with more training data.
If there are more instances per annotation, the diversity in the related subtrees will usually increase, which reduces the number of patterns that can be found and thus reduces the training time.

\section{Conclusions and future work}
\label{subsec:feedbackpredictionconclusion}
We presented ECHO as a predictive method to assist human reviewers in giving feedback when reviewing students submissions to a programming exercise by reusing annotations.
Improving the reuse of annotations can both save time and improve the consistency with which feedback is given.
The latter in itself might further improve the accuracy of predictions if the strategy is applied during the review process.

ECHO has already shown promising results.
We have validated the framework both by predicting automated linting annotations to establish a baseline, and by predicting annotations from human reviewers.
The method has about the same prediction accuracy for machine (Pylint) and human annotations.
Thus, we can answer both our research questions in an affirmative way, meaning that the reuse of feedback previously given by a human reviewer on a particular line of a new submission can be predicted with high accuracy (RQ1), and that this can be done fast enough to assist human reviewers in future reviews (RQ2).

Having ECHO at hand immediately raises some opportunities fo follow-up work.
Currently, the proposed model is reactive: we suggest a ranking of the most likely annotations when a reviewer wants to add an annotation to a particular line of a submission.
By introducing a confidence score, we could check beforehand whether we have a confident match for each line, and then immediately propose these suggestions to the reviewer.
Whether or not a reviewer accepts these suggestions could then also be used as an input to the model.
This could also have an additional benefit by helping reviewers to be more consistent in where and when they place annotations.

Annotations that don't lend themselves well to prediction also need further investigation.
The context used could be expanded, although the important caveat here is that the method still needs to maintain sufficient performance.
We could also consider applying some of the source code pattern mining techniques proposed by~\citet{phamMiningPatternsSource2019} to achieve further speed improvements.
This could help with the outliers seen in the timing data.
Another important aspect that was explicitly outside of the scope of this manuscript was the integration of ECHO into a learning platform and user testing.

Of course, alternative methods could also be considered.
One cannot overlook the rise of Large Language Models (LLMs) and the way in which they could contribute to this problem.
LLMs can generate feedback for students based on their submitted solution and a well-chosen system prompt.
Fine-tuning of an LLM with feedback already given is another possibility.
Future applications could also combine user generated and LLM generated feedback, showing human reviewers the source of the feedback during their reviews.

\bibliographystyle{apalike}
\bibliography{sn-bibliography}

\section*{Statements and Declarations}

\subsection*{Funding}

We are grateful for the financial support of Ghent University (Belgium) and the Flemish Government (Belgium, Voorsprongfonds) through numerous innovation in education grants. Part of this work was also supported by the Research Foundation - Flanders (FWO) for ELIXIR Belgium (I002819N and I000323N).

\subsection*{Competing interests}

The authors have no relevant financial or non-financial interests to disclose.

\end{document}